\documentclass{article}
\usepackage{graphicx}

\usepackage{microtype}
\usepackage{subfigure}
\usepackage{booktabs} 

\usepackage{hyperref}

\usepackage[accepted]{icml2024}

\usepackage{mathtools}
\usepackage{amsthm}

\usepackage[capitalize,noabbrev]{cleveref}

\theoremstyle{plain}

\theoremstyle{definition}

\theoremstyle{remark}

\usepackage[textsize=tiny]{todonotes}

\icmltitlerunning{Wavelet-GPT: Wavelet Inspired Large Language Models}

\begin{document}

\twocolumn[
\icmltitle{Wavelet GPT - Wavelet Inspired Large Language Models}



\icmlsetsymbol{equal}{*}

\begin{icmlauthorlist}
\icmlauthor{Prateek Verma}{yyy}
\end{icmlauthorlist}

\icmlaffiliation{yyy}{Department of Computer Science, Stanford Natural Language Processing Group, Stanford University, Stanford, CA 94305. This work was done while Prateek Verma was with the Stanford Natural Processing Group in 2023}\icmlcorrespondingauthor{Prateek Verma}{prateekv@stanford.edu} 
\icmlkeywords{Machine Learning, ICML}

\vskip 0.3in
]



\printAffiliationsAndNotice{}  
\begin{abstract}
Large Language Models (LLMs) have ushered in a new wave of artificial intelligence advancements impacting every scientific field and discipline. We live in a world where most of the data around us, e.g., text, audio, and music, has a multi-scale structure. This paper infuses LLMs with a traditional signal processing idea, namely wavelets, during pre-training to take advantage of the structure. Without adding \textbf{any extra parameters} to a GPT-style LLM architecture in an academic setup, we achieve the same pre-training performance almost twice as fast in text, audio, and images. This is done by imposing a structure on intermediate embeddings. When trained for the same number of training steps, we achieve significant gains in performance, which is comparable to pre-training a larger neural architecture. Further, we show this extends to the Long Range Arena benchmark and several input representations such as characters, BPE tokens, bytes, waveform, math expression, and image pixels.  Our architecture allows every next token prediction access to intermediate embeddings at different temporal resolutions in every decoder block. We hope this will pave the way for incorporating multi-rate signal processing into pre-training instead of going after scale. 
\end{abstract}
\vspace{-0.6cm}

\section{Introduction and Related Work} 
LLMs have ushered in a super-renaissance of AI advancements and are touching every scientific and engineering discipline. At the heart of this is the Transformer architecture \citep{vaswani2017attention}, initially proposed for machine translation. Transformer architecture became the backbone of GPT (Generative Pretrained Transformer) language models \citep{brown2020language} first proposed by Open-AI. Modern LLMs are trained on a straightforward objective: To predict the next token given the previous context, preserving the causality. This not only works for language but also for robotics \citep{brohan2022rt, brohan2023rt}, protein sequences \citep{madani2020progen}, raw audio waveforms\citep{verma2021generative}, acoustic/music tokens \citep{huang2018music, verma2020framework,borsos2023audiolm}, videos \citep{yan2021videogpt} etc. This simple recipe of tokenization/creating an embedding and feeding it to transformers also has given rise to non-causal architectures such as BERT\citep{devlin2018bert}, Vision Transformers \citep{dosovitskiy2020image}, Audio Transformers \citep{verma2021audio} and Video Transformers \citep{selva2023video}. With increased performance by scale, LLMs are reaching hundreds of billions to trillions of parameters \citep{brown2020language,fedus2022switch}. 

Recent concerns suggest AI research is shifting from academia to industry, according to a Washington Post article by \citep{ai-pricing-out}. This work aims to enhance LLM capabilities to match those of larger architectures or achieve equivalent performance in fewer training steps. Knowledge distillation \citep{hinton2015distilling}, uses a larger model to guide a smaller one. \citep{gu2024minillm} used KL divergence to enhance next-token prediction from teacher model feedback model rather than training the smaller one from scratch. Model pruning \citep{sun2023simple} removes weights to match the same performance as a large model like LLAMA \citep{touvron2023llama}, with fewer compute flops during inference, still relying on a larger model. \cite{dettmers2024qlora} focus on improving inference or fine-tuning existing models. Unlike distillation and pruning our approach focuses on improving performance during pre-training from scratch. \citep{nawrot2021hierarchical}, proposed hierarchical transformers using upsampling-downsampling operations achieve results comparable to those of Transformers but with more efficient computation. Clockwork RNN \citep{koutnik2014clockwork} improves long-context modelling by splitting RNN neurons into modules that update at different clock rates. Only a few modules activate at each time step. Our approach modifies intermediate embeddings with simple tweaks without using separate learning modules or varying update rates. 

Tinkering with the intermediate embeddings: \cite{tamkin2020language} proposed hand-tuned filters on the Discrete Cosine Transform- DCT \citep{ahmed1974discrete} of the latent space for different NLP tasks for non-causal BERT \citep{devlin2018bert}. Computing DCT over context length makes it not applicable for causal applications such as language modelling. There has been work on applying ideas from signal processing-like methods to BERT-like non-causal architectures. We discuss two here, FNet and WavSPA. They focus on improving attention block, which differs from our work on GPT, which retains a vanilla attention layer. FNet proposed by \cite{lee-thorp-etal-2022-fnet} removes the costly attention mechanism, replacing it with a 2-D FFT block. This operation is non-causal as it looks into future tokens for computing 2-D  FFT. WavSpA \citep{zhuang2024wavspa} carries attention mechanism in the wavelet space. The input sequences are transformed into wavelet space, and the attention mechanism is carried out and then reconstructed. However, computing wavelet transform is non-causal, making them non-applicable for GPT-based LLMs as they look at the entire sequence length (Fig 1 \citep{zhuang2024wavspa}). 

Our work is inspired by neuroscience, which provides evidence that human brain learns multi-scale representations for language at multiple time scales \citep{caucheteux2023evidence} instead of fixed-resolution representations. We impose multi-scale representation onto every intermediate decoder embedding at different dimensions. To the best of our knowledge, the paper's contributions are: 1) We propose the first instance of incorporating wavelets into LLM pre-training. We add multi-scale filters onto each of the intermediate embeddings of decoder layers using the Haar/learnable wavelet pipeline. This allows every next token prediction access to multi-scale intermediate embeddings instead of being fixed-resolution in every decoder layer representation. 2) We show speedups in pre-training of GPT, like transformer-based LLM in the range of 40-60\%, with adding a multi-scale structure. With same training steps, the model gives a performance boost akin to adding several layers.

 \section{Dataset} We use four open-source datasets from four domains for next token prediction: natural language, symbolic music, speech tokens, and raw audio waveform. For text, we choose text-8 \citep{mikolov2012subword}. We choose this over other datasets as i)it is a famous and widely cited character-level language modelling dataset, and ii) it uses a simple vocabulary (space + 26 lowercase characters) to detach the effects of various tokenizers. It has 100M characters with split training split as given by \cite{al2019character}. For raw audio, the goal is to predict the next sample given the context. We use the YouTube-Mix-8 dataset for long-context modeling \citep{goel2022s,verma2022goodbye}. Our vocabulary size is 256, with a sampling rate 16KHz as input is 8-bit. We use a third dataset, MAESTRO \citep{hawthorne2018enabling}, containing over 1000 MIDI files of classical music pieces with a tokenizer proposed by \cite{huang2018music}, which converts MIDI tracks into discrete tokens with a vocabulary size of 388. Finally, we use 1000 hours of LibriSpeech dataset and a widely used ENCODEC \cite{defossez2022high} tokenizer in a setup similar to VALL-E \cite{wang2023neural} to model acoustic tokens \footnote{The goal in this work, is how well we can model the coarsest tokens, as errors in modelling the coarser tokens will lead to the finer tokens being modelled incorrectly as they are conditioned on the coarsest token, thus affecting model performance}. The goal in all four modalities is not to chase state-of-the-art pre-training performance, as \textit{this paper was written in an academic setting with very few computational resources}. We show how the model performs in pre-training instead of post-training, as the goal is to build better foundational architectures with the same parameters, pushing the capabilities of smaller decoder architectures.
\begin{figure}[ht]
\begin{center}
\centerline{\includegraphics[width=\columnwidth,height=4.1cm]{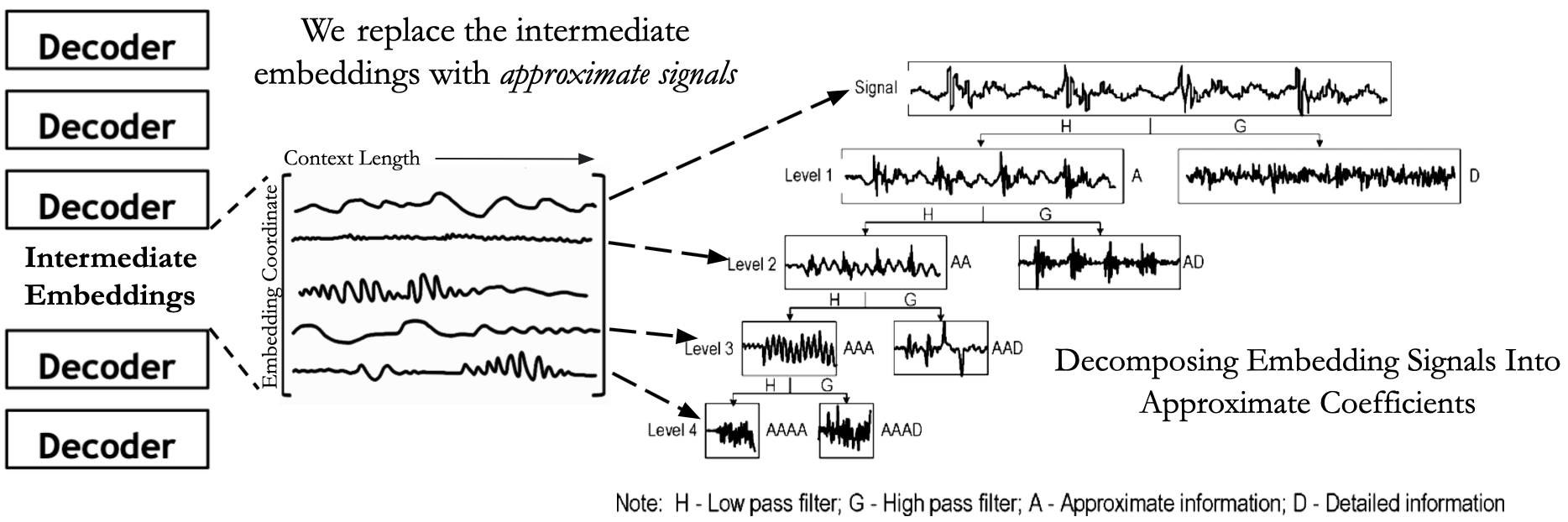}}
\caption{Manipulating signals between GPT decoder blocks by computing 1-D causal discrete haar wavelet transform/learnable approximation at different levels capturing multi-scale structure for each signal. (Right) From \cite{gao2006non} explaining non-stationary signal processing for signals. Leftmost route of approximate coefficients to model coarsest to finest scales.}
\label{icml-historical}
\end{center}
\vspace{-0.5cm}
\end{figure}

\begin{figure}[t]
\begin{center}
\centerline{\includegraphics[width=\columnwidth, height=5cm]{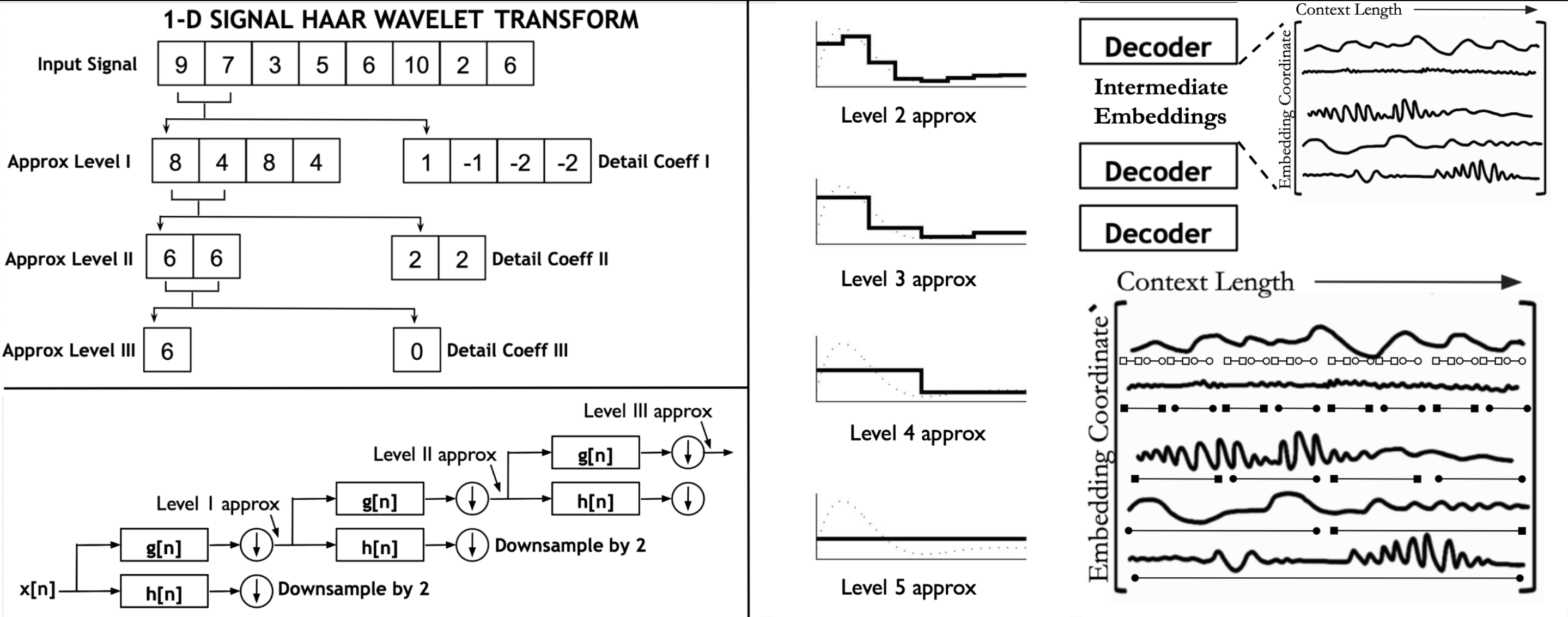}}
\caption{ (Bottom L): A 3-level filter bank tree generates signals at different resolutions. Approximate coefficients are computed by applying a wavelet's impulse response \& recursively down-sampling. (Top L): Approximate and detailed coefficients are iteratively calculated via first-order averages/differences and down-sampling until a single scalar represents the signal. (R): For a 32-length signal, Haar wavelet captures coarsest to finest approximations and is redrawn from \citep{wavelet-tutorial}. Embeddings evolve at different rates via causal wavelet approximation, with coarse (level 5) and fine (level 2) resolutions, embedding multi-scale information into decoder layers for every token.}
\label{icml-historical}
\end{center}
\vspace{-0.8cm}
\end{figure}
\section{Methodology} This section will describe the approach to incorporating wavelet inspired computation into transformer-based Large Language Models while retaining causality. The ideas described here are generic as we tinker intermediate embeddings. Thus theycan be easily extrapolated to non-Transformer architectures, e.g. state space model. For any GPT signal, we compute a version of the discrete wavelet transform and incorporate it back into the signal. Let \( x_{(i)}^{l} \) be the output of the \( l^{th} \) decoder layer, representing the activation along the \( i^{th} \) coordinate, with a dimension equal to the context length \( L \) of the transformer-based GPT model. In the original GPT architecture with \( N+1 \) layers and embedding dimension \( E \), we obtain \( N \cdot E \) signals of length \( L \) from intermediate embeddings between decoder blocks, where \( E \) ranges from \( [0-128) \) dimensions. For any signal \( x[n] \), the discrete wavelet transform resembles passing the signal through filters of varying resolutions, as illustrated in Figure 2. We will use the Haar wavelet, a family of square-shaped functions this paper obtained from a mother wavelet via scaling and shifting operations.
Given a mother wavelet function $\psi$, the child wavelets as $\psi_{j, k}[n]$, where $j$ is the scaling factor and $k$ is the shift factor. \begin{equation}
    \psi_{j, k}[n]=\frac{1}{\sqrt{2^j}} \psi\left(\frac{n-k 2^j}{2^j}\right) \end{equation} These signals are shifted and scaled to capture information at various time scales, with \( n \) representing time or the context length. This concept resembles the diagram in Figure 1, which illustrates capturing different signals in the intermediate layers of Transformer decoders at various resolutions. Discrete wavelet transform, which passes any signal through filters and downsampling operations. This process, shown in Figure 2, is similar to a convolutional neural network (CNN) like ResNet \citep{he2016deep}, featuring learned convolutional filters analogous to \( h[n] \) and \( g[n] \), along with downsampling, such as max pooling. In convolutional architectures, we follow one branch of Figure 2, recursively taking the output of filters and downsampling. This similarity contributed to popularity of wavelets in the 1990/2000s for image understanding, reflecting parallels with convolutional architectures \citep{huang2008wavelet, kingsbury1998wavelet}. For Haar wavelets, this is passing the signal through low-pass and high-pass filters corresponding to the kernels \( g[n] \) and \( h[n] \). The Haar wavelet transform averages and computes differences, with impulse responses \( g[n] = \left[ \frac{1}{2}, \frac{1}{2} \right] \) and \( h[n] = \left[ \frac{1}{2}, -\frac{1}{2} \right] \). Figure 2 provides a detailed explanation of the discrete wavelet transform. For a 1-D signal \( x[n] \) of length \( L \), we get level 1 coefficients by filters \( g[n] \) and \( h[n] \), followed by downsampling. Thus, the approximation coefficients \( y_{\text{approx}} \) and \( y_{\text{detail}} \) result from a LTI system defined by convolution followed by downsampling by two (Equation 2). This is seen in Algorithm 1 with \( \text{type}\in\{\text{approx}, \text{detail}\} \) and 
\( f_{\text{approx}} = g \), \( f_{\text{detail}} = h \). \begin{equation}y_{\text{type}}[n] = \sum_{k=-\infty}^{\infty} x[k] \, f_{\text{type}}[2n - k]\end{equation}To obtain multi-scale representations of the original signal, the operation for \( x[n] \) is recursively applied to \( y_{a} \) (approx) to derive level 2 wavelet coefficients \( y^{2}_{a} \) and \( y^{2}_{d} \) (detail). Here, \( x[n] \) represents intermediate signals across the context length at each decoder block output in the LLM. The approximate coefficients \( y_a \) and \( y_d \), along with their decompositions \( \{y_{a}, y_{d}, y^{2}_{a}, y^{3}_{a}, y^{4}_{a}, \ldots\} \), are used for further processing. Notably, \( y^{2}_{a}, y^{3}_{a}, y^{4}_{a} \) have lengths reduced by factors of \( 2, 4, 8, \ldots \). The Haar wavelet transform averages adjacent samples while preserving causality by averaging current and past samples. Higher-order coefficients capture averages over larger context lengths, as shown in Figure 2. We can continue until only a single scalar value remains, representing the mean of the signal. The Haar wavelet transform computes averages and differences to create a multi-resolution representation, capturing low and high frequencies at different resolutions. Figure 2 illustrates the same signal captured at coarser and finer representations using Haar wavelets, applied to intermediate embeddings, allowing each next token prediction access to these representations. For the case of learnable wavelet kernels, we create a multi-resolution representation by varying the kernel size (Algorithm 1) to allow the LLM to learn the optimal kernels optimized for the next token prediction.

\subsection{Connecting wavelets and LLM embeddings}
In many signal processing applications, first order detailed along with approximate coefficients captures signals at various levels. We do the same with signals from intermediate transformer embeddings across tokens. Our premise is that real-world data is structured: text ranges from letters to words, sentences, and topics, symbolic music ranges from notes, motifs and pieces with speech from phonemes, morphemes to syllables to phrases. With Haar wavelet, this can be approximated as a simple averaging operation. For the learnable version, the weights of the kernel for the multi-scale version are optimized to predict the next token. Continuing with the approximate coefficients will eventually yield a single scalar, the average of the entire signal in the case of the Haar wavelet. Several methods can be employed to match the original signal's sequence length from the approximation coefficients, including up-sampling. We refer to the signal approximated at a specific level with the same length as the "approximate signal" distinguishing it from shorter approximate coefficients. In Figure 2 (R), to obtain the signal approximation at various levels matching the original input signal \(x[n]\), we apply the wavelet kernel by multiplying the approximate coefficients with the kernel for that level (e.g., \([1,1]\), \([1,1,1,1]\), etc.). This piecewise constant function is shown in Figure 2. LLM embedding coordinates define unique resolution kernels, each corresponding to a specific scale of data. The reconstructed signal \(x_{\text{recon}}[n]\), a method to derive the *approximate signal*, is computed from wavelet coefficients \(c_{j}\) at level \(j\) as: \begin{equation}
x^{j}_{\text{recon}}[n] = \sum_{k} c_k \cdot \psi_{j, k}[n]
\end{equation} 
\vspace{-0.9mm}
Equation 3 requires storing child wavelets at various approximations, complicating the process and rendering it non-causal as computing $c_{k}$ considers the entire signal. Due to the dependence of $c_{k}$ on future information, we cannot use this to reconstruct the signal from its approximate coefficients. To adapt this for LLM, we simplify the computation of the *approximate signal* in a differentiable manner using a variant from Equation 3 in both multi-resolution learnable/non-learnable kernel settings. For the Haar wavelet, we compute an average of the input signal with varying kernel lengths, increasing the length until it approximates the entire signal. The kernel length determines the level of signal approximation. LLMs operate under a causality assumption, modifying the signal at a location using prior samples within the kernel length. We zero-pad the signal to the left when the window length is shorter than the kernel. Wavelet transform at different levels gives several versions of the signal at different resolutions, which can mess up the structure of the intermediate embeddings. To address this, we create different resolutions for signal approximations parameterized by embedding dimension. In Section 4.4, we make these kernels learnable, allowing the architecture to maintain multi-scale operation (Equation 3), with learnable weights with $x_{\text{recon}}[n]$ now being learned with resolution parameterized by latent dimension.

\begin{figure}[t]
\begin{center} \centerline{\includegraphics[width=0.9\columnwidth,height=6.1cm]{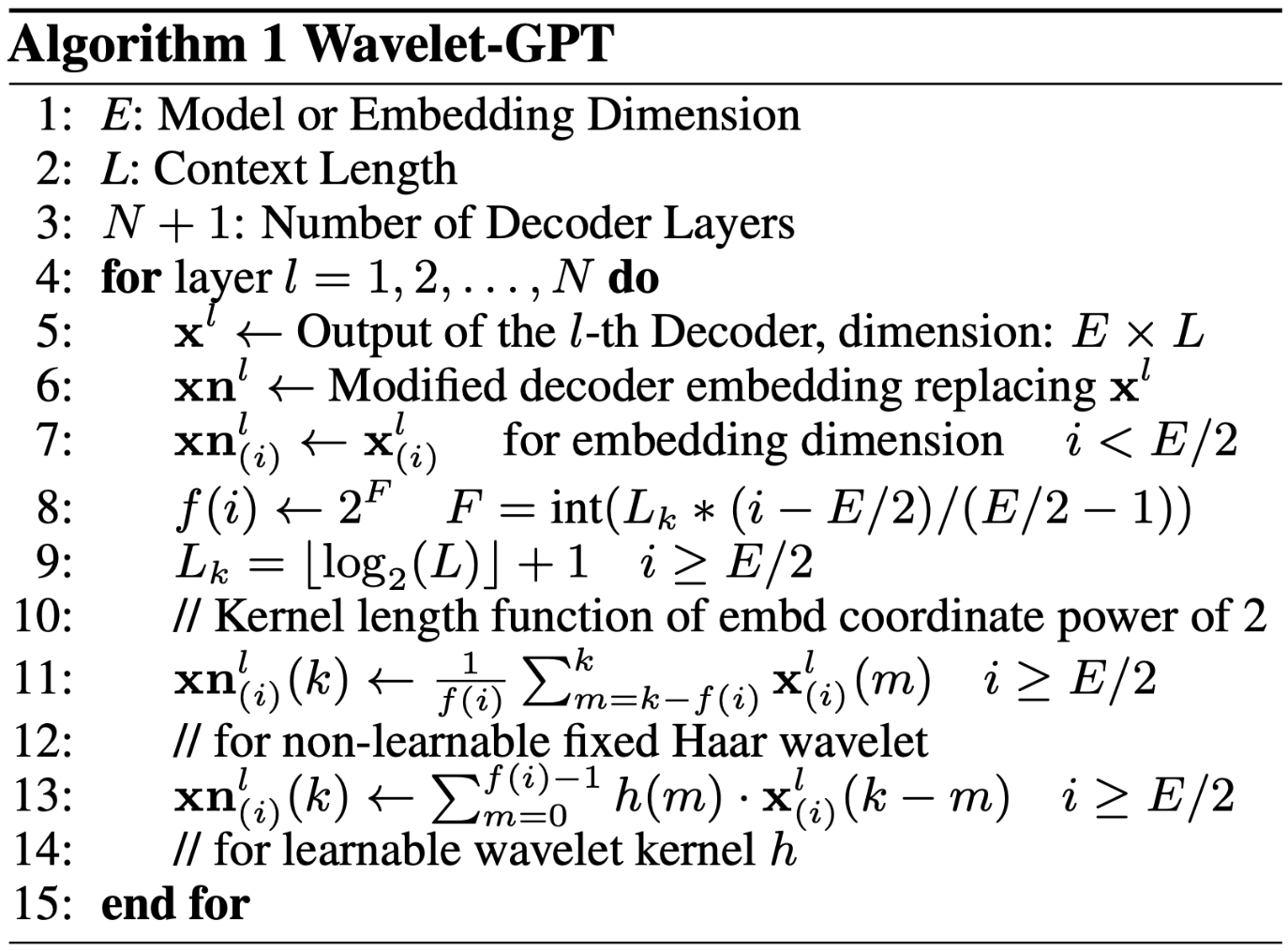}} \vspace{-0.9cm} \end{center}  \end{figure}

\begin{figure*}[t]
\begin{center} \centerline{\includegraphics[width=\textwidth, height=6.2cm]{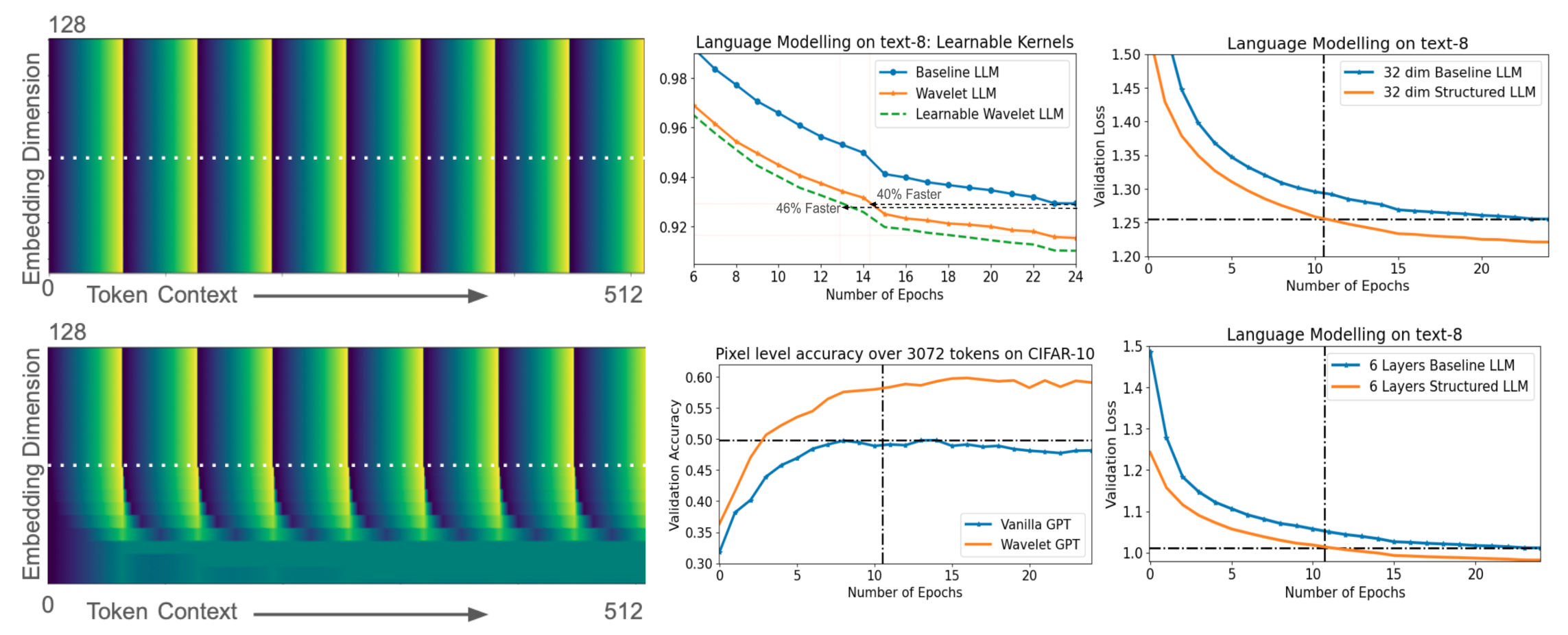}} \caption{(Left) Toy example showing embeddings before/after imposing multi-rate structure. Different embedding dimensions advance at distinct rates while maintaining causality, as seen from patterns dispersing from dimension 64 to 0. (Right) Validation loss during pre-training on text-8 with learnable multi-scale structure achieving comparable performance nearly twice as fast/performance boost akin to adding additional decoder layers. Our architecture's performance on text-8 with a 32-dim model matches the speedup similar to that seen for 128-dim and shallower models. LRA image benchmark, a 10\% performance increase without adding any parameters} \vspace{-0.9cm} \label{icml-historical} \end{center}  \end{figure*}
\subsection{Wavelet Coefficients by Embedding Coordinates}
One option is to compute the \textit{approximate signals} for each coordinate signal $x_{(i)}^{l}$ across all decoder layers at levels I to IX. For a context length of 512, this would require nine additional signals with resolutions of 512, 256, 128, 64, 32, 16, 8, 4, and 2, significantly increasing complexity and necessitating major modifications to our GPT model. We propose a novel solution to address this. Instead of computing all levels of \textit{approximate signals} for every intermediate embedding dimension, we parameterize the level by the embedding dimension index. We want to steer the embeddings only a little into the inductive biases we impose to avoid too much tinkering with what they learn. Transformers have been wildly successful without incorporating any inductive biases. Ideally, we want the best of both worlds, nudging intermediate GPT embeddings in only half of the dimensions. We adjust intermediate GPT embeddings in only half the dimensions. Embeddings from $0$ to $E/2$ (coordinates 0 to 64 when $E=128$) remain unchanged. For the rest, we apply processing based on their index $i$. If $x^{l}{(i)}$ is an intermediate embedding after the $l^{th}$ decoder layer along the $i^{th}$ dimension, the modified signal $xn^{l}{(i)}$ equals $x^{l}_{(i)}$ for $i \in [0,E/2]$. For $i > E/2$, we impose structure using an approximate signal calculated from wavelet coefficients corresponding to the index $i$. We use a mapping function $f$ that takes coordinate $i$ (ranging from $E/2$ to $E$) and returns the kernel size corresponding to approximation levels from $I$ to $IX$. The linear function gradually increases from level $I$ (kernel size two at $i=E/2$) to level $IX$ (kernel size 512 at $i=E$, or the coarsest representation i.e., a scalar). 

Now, let us find out how we compute the modified new signal $xn^{l}_{(i)}$ that replaces the original intermediate Transformer embeddings $x^{l}_{(i)}$. $f(i)$ is the kernel size for the coordinate $i$. The modified signal is either kept the same or modified as $xn_{(i)}^l(k)=\frac{1}{f(i)}\sum_{m=k-f(i)}^k x_{(i)}^l(m)$ as seen in Algorithm 1. For cases where \( k - f(i) < 0 \), we zero-pad the signal to ensure valid average/kernel computation. Specifically, for the Haar wavelet, the modified signal acts as a causal moving average filter with finite length, averaging the embedding signal along the \( i^{\text{th}} \) coordinate with a kernel size determined by \( f(i) \). This operation does not introduce new parameters or maintain causality in LLMs to prevent future token leakage, as seen in Equation 4. In Algorithm 1, each value of the modified signal at token \( k \) is computed using a convolution with a learned kernel \( h(.) \) and variable length \( f(i) \), parameterized by the embedding coordinate dimension \( i \). Each kernel is learned independently for every signal.  

\subsection{Imposing Structure: Toy Example} In Figure 3, we illustrate a toy example of how we impose structure onto decoder Transformer embeddings. The left side shows eight variations along the token dimension, with onset/sudden bursts at token indices 32, 64, etc., decreasing to zero before rising again. As discussed in the introduction, datasets inherently possess a hierarchical structure, which we capture by imposing intermediate Transformer embeddings at each layer. In this example, we retain embeddings at the original resolution for half the dimensions (split by a white line). For the other half, we gradually increase the kernel length across the context and compute the average causally. The final embedding dimension averages over the token dimension with a kernel size equal to the context length (zero-padding if necessary). This creates highways, allowing embeddings to move at different rates: the coordinates from $E/2$ to $ E$ move at the Transformer's original speed, while those from 0 to $E/2$ transition from faster to slower movement. This approach enables the attention mechanism to utilize multi-scale features at varying rates across all layers and tokens, as explored in the next section. Further, this multi-scale structure can be made learnable, driven by just the next token prediction. 
\section{Experiments} The main aim of these experiments is to show that the pre-training performance of the models across four modalities improves with/without doing intermediate modifications on embeddings inspired by wavelets. We also benchmark on LRA tasks. We propose a shrunk-down GPT baseline architecture that has the same topology. We do not compare against larger architectures, as this paper focuses on pre-training from scratch, and was written in with access to limited computational resources in academia. We evaluate pre-training performance with and without wavelet-inspired blocks. We only report how well our generative model does for pretraining by quantifying the likelihood scores similar to papers such as Mega-Byte \cite{yu2023megabyte} and Music Transformer \cite{huang2018music} that only report NLL scores in the entire paper. We also validate our method across various modalities such as text, audio and music. Further, we benchmark it on various input representations such as raw waveform, MIDI tokens, acoustic tokens, text bytes, characters, and BPE tokens, in addition to math expressions.  Our experiments, based on the GPT-2 architecture, have 10 Transformer decoder layers with a context length of 512, trained from scratch. Each modality shares the same architecture, using an embedding dimension of 128, a feed-forward dimension of 512, and 8 attention heads. The final decoder outputs a dense layer of 2048 neurons, followed by a layer matching the vocabulary size \footnote{Vocab of 27 for text8, 256 for raw waveform \citep{goel2022s,verma2022goodbye}, 388 for symbolic music, and 1024 for ENCODEC speech tokens, 256 for 8-bit raw pixels, 50257 for BPE tokens}. 
\begin{figure*}
\begin{center}
\centerline{\includegraphics[width=0.9\textwidth,height=3.5cm]{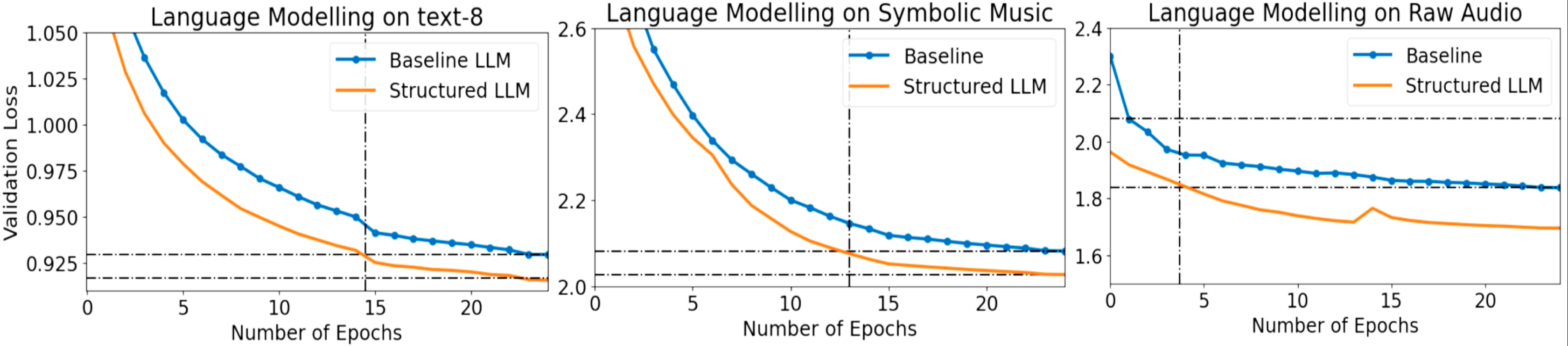}}
\caption{Results for natural language, symbolic music, and raw audio. We perform faster than baseline, almost twice as fast on shrunk-down GPT. We see substantial gains in pre-training performance for the same epochs, equivalent to a much larger architecture. The black vertical line denotes the epoch at which our architecture achieves the same performance as our baseline architecture.}
\label{icml-historical}
\end{center}
\vspace{-0.8cm}
\end{figure*}
 Baseline models consist of standard Transformer decoder blocks without modified embeddings. We retain half \footnote{The choice of half is a hyper-parameter and it is difficult to optimize for every modality and input represenation due to computational resource constraints. If the optimal split is not half, it will only improve the already strong results} of the embedding coordinates for our proposed architecture and impose either a fixed or learnable multi-scale structure on the other half for all intermediate layers.  All models were trained from scratch in TensorFlow \cite{abadi2016tensorflow} for 25 epochs, starting with a learning rate of 3e-4, decreasing to 1e-5 when loss plateaued. Each model utilized 1M training points, totalling 500 million tokens, randomly cropped from the dataset. We measured performance using negative log-likelihood loss, as this method improves the core architecture of the transformer-based GPT - helping achieve the objective we want to achieve: predict the next token correctly. Since we are operating on intermediate embeddings, our work can hopefully generalize to setups with structured data similar to text, raw audio, and symbolic music, where one can go from a fine-grained structure to a coarse structure. We impose a multi-scale structure that allows the attention mechanism to learn dependencies across embeddings and inject some information that can capture coarse and fine-grained structures into embedding coordinates while maintaining causality. 

\subsection{Performance on modalities} We compared the performance of our baseline architecture across three modalities : text, symbolic music, and audio waveform with and without wavelet-based intermediate operations. Results showed significant performance improvements in all modalities with the same number of training steps. To illustrate, a 0.04 decrease in validation loss is comparable to going from a 16 to a 64-layer model on text-8 dataset \citep{text8-paper}. As shown in Figure 4, our modified GPT architecture achieves this loss nearly twice as quickly in training steps as the original model, showing that GPT-like architecture can take advantage of the structure we imposed on half of the embedding dimensions. This speedup, i.e., the number of epochs/steps taken to achieve the same performance (SP: same performance epoch), is even smaller for raw audio due to the quasi-stationary nature of audio signals at smaller time scales (20-30 ms for harmonic sounds). For a sampling rate of 16KHz, a context length of 512 would correspond to 32ms, which may be one of the reasons that some of the coordinates nail down the contents of the context in fewer coordinates onto which we impose structure. The convergence is significantly faster for the raw waveform LLM setup and achieves nearly twice the speed of text-8 and symbolic music. 
We also ran benchmarks on LibriSpeech corpus about 1000 hours of speech and acoustic tokens, further strengthening our method is generic to handle several types and modalities of tokens. We run our method on audio classification with Audio Transformer over 200 categories of audio for FSD-50K benchmarks \cite{fonseca2020fsd50k}. We get a performance boost and speedups, thereby showcasing the ubiquity of our proposed method for generative modelling and classification as well. We also compare the absolute clock run times of our modifications in both learnable/non-learnable setups. Table 1 reports the time to complete one epoch relative to our baseline architecture. Our method is computationally inexpensive, as it only involves fixed kernel multiplication or learning a single filter convolutional kernel with variable context lengths along different coordinates. 

\begin{figure*}[ht]
\centering
\begin{minipage}{0.7\textwidth}
  \centering
  \begin{tabular}{|c|c|c|c|c|c|}
    \hline
    Modality & Baseline & Proposed & SPE & SpeedUp & Rel. GPU Hrs \\\hline
    Text-8  & 0.93  & 0.92 & 14.5 & 42\% & 1.013\\
    Raw Audio & 1.84 & 1.70 & 3.7 & 85\% & 1.042\\
    Symbolic Music  & 2.08  & 2.02 & 13  & 48\% & 1.059 \\
    Text-8 (L)  & 0.93  & 0.91 & 12.9  & 48\% & 1.094\\
    Wiki-103 (L) & 4.11 & 4.05 & 9.5 & 62\% & 1.130\\
    LibriSpeech (L) & 2.43 & 2.40 & 9.2  & 63\% & 1.110 \\
    FSD-50K & 40.6\% & 42.8\% & 32 & 65\% & 1.037 \\ \hline
  \end{tabular}
\end{minipage}%
\hfill
\begin{minipage}{0.3\textwidth}
  \centering
  \includegraphics[width=\columnwidth,height=4cm]{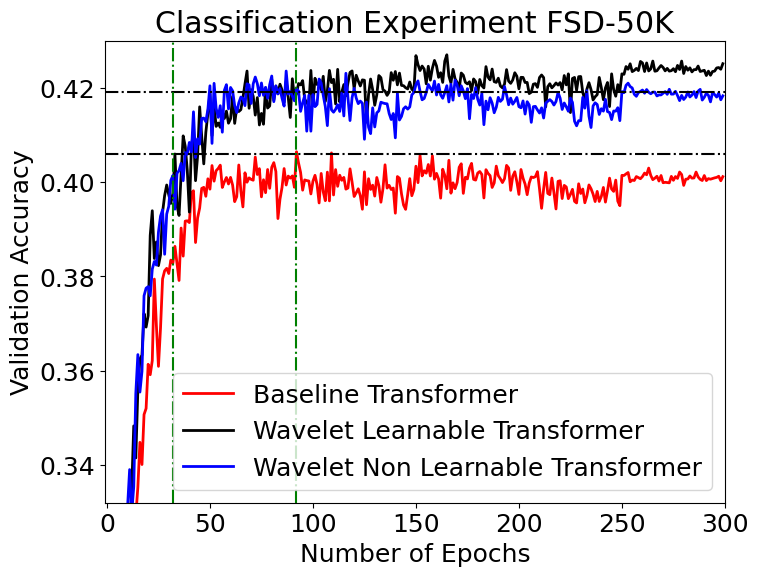}
\end{minipage}
\vspace{-0.4cm}
\caption[Comparison of Wavelet-Based Architecture Performance]{Comparison of the negative-log likelihood (NLL) scores for our architecture across three modalities, with and without wavelet-based fixed/learnable (L) structure. (Left) Table shows the NLL scores and speedup, with Same Performance Epoch (SPE) with baseline as 25 epochs, relative GPU hours. (R) FSD-50K Audio Transformer top-5 accuracy results. Vertical green lines indicate the highest accuracy achieved and the point where the same accuracy is reached 60\% faster using our proposed method, with no extra parameters.}
\label{fig:combined-results}
\vspace{-0.5cm}
\end{figure*}

\subsection{Making multi-scale kernels learnable} We allow each kernel to be learnable. In the previous section, we defined the shape of the kernel and computed approximate signals of intermediate layer activations across all layers, with different resolutions occurring at different embedding dimensions to mimic a causal version of the wavelet transform. Now, we allow each kernel of length $L$ at a particular level to be learnable for computing the \textit{approximate signal} for various resolutions, yet another way to compute it. By making the computation of approximate signal learnable, the model can learn how to weight every decoder layer dimension instead of putting a fixed kernel, e.g. exponentially weighted average. This as can be seen, Algorithm 1 only allows 20k or 0.2 \%  extra parameters to our base decoder architecture. We run the experiments both for learnable kernels and fixed kernels (three for each case) due to scarcity of resources. Intuitively there will exist an optimal kernel whose shape is best optimized for the next token prediction for a specific modality of interest than simple Haar wavelet. This further improves our performance from 42\% to 48\% faster speedup to get a similar baseline performance, seen in Figure 4, carried out on the text-8. We also benchmark on Wiki-103 to demonstrate that our method works with the GPT-2 tokenizer giving even larger gains. As shown in Figure 5, we match the performance of a 10-layer architecture at more than twice the speed. In addition to faster convergence, we see a 3.6-point improvement in perplexity scores over the baseline model for Wiki-103. Section 4.4 shows we scale with model size and depth showing promise for improvements in larger LLM architectures. 

\subsection{Ablation on Depth And Model Dimension} The aim for these experiments was to see if our model scales with depth of the Transformer and the model dimension. We explore two architecture variants on text-8: (i) reducing the model dimension from 128 to 32 and (ii) reducing the number of layers. The model with 32-dimensional, 10 decoder layers (eight heads) achieves baseline performance in around ten epochs and runs nearly twice as fast (Figure 4). For the second experiment, we retain the architecture from Table 1 but reduce the Transformer decoder to six layers while keeping other parameters unchanged (feed-forward dimension four times the model dimension, eight heads). With Haar-inspired modifications, the model matches baseline performance twice as fast, consistent with the results reported in Section 4.1. While it is difficult to scale the architectures beyond a certain depth and model dimension in academic setups: we believe that by seeing the effect of model dimension and depth holding, we are confident that the findings will extrapolate for much larger/deeper models.

\subsection{Audio Classification Benchmark} We explore the strength of our method for a typical audio classification on a standard audio classification benchmark FSD-50K. The goal is to identify the sound categories in 1s of audio correctly. We use a transformer-based architecture similar to Audio Transformer as our baseline model. It consists of 128 convolutional filters with a length of 200 learned over 25ms of audio sampled at 16KHz, yielding patches of 400-length audio samples. The convolutional filter output is then max-pooled across the 25ms window to give a single vector of length 128, i.e. the number of convolutional filters being fed to a Transformer stack of 6 layers with model dimension as 64 similar to \cite{verma2023content,verma2021audio}. We report on the top-5 \% accuracy and relative gain in mAP scores for our proposed architecture with learnable/non-learnable kernels. We see that we get a performance boost of about 2\% and a faster convergence of more than 60\%, as reported in Figure 5, for the learnable/fixed kernels and baseline. \subsection{Similarities and Differences With EMA} We compare Exponential Moving Averages (EMA) on intermediate signals. Unlike the Haar wavelet, which takes fixed window weights, which takes the mean of the signal in the window, EMA uses an exponential kernel. Let the signal $x_{i}^{l}(t)$, after the $l^{th}$ layer, be of length equal to context length, with $t$ being the token index from 0 to $L$ at embedding dimension $i$. The modified signal $s_t$ is: $ s_0 = x_i^l(0) \quad s_t = \alpha x_i^l(t) + (1-\alpha) s_{t-1}$ where $\alpha$, the decay factor, satisfies $0 < \alpha < 1$. Unlike an EMA, our method captures multi-scale information using a finite kernel with zero weights outside a specified length. In text-8 experiments, we applied EMA on half of the embedding dimensions, with $\alpha$ linearly varying between 0 and 1 for dimensions 64 to 128 mimicing our approach with a classical approach. This under-performed compared to our baseline, with an NLL score of 0.94, while our baseline and proposed method achieved scores of 0.93, 0.92, and 0.91 for non-learnable and learnable cases, respectively. Our method provides a simple, signal processing-based scheme that optimizes weights across multiple resolutions driven by next-token prediction and outperforms EMA. Depending on $\alpha$, the EMA filter produces an exponential kernel while we maintain a constant kernel or allow weights learned from scratch optimized for the next token prediction. Further, EMA is an Infinite-Impulse Response (IIR) filter, whereas the Haar wavelet-based kernel is a Finite Impulse Response (FIR) filter. Consequently, for each value update, the contributions from previous samples never reach zero. These can accumulate significantly at longer context lengths for certain $\alpha$. The recursive, non-learnable nature of the EMA IIR filter ensures some contribution from all embeddings, which explains performance degradation. In contrast, our method uses zero weights outside the kernel length, capturing multi-scale information.

\begin{table}[htbp]
\vspace{-0.3cm}
\centering
\caption{Performance on LRA tasks (\cite{Tay2020b}) as reported in \cite{liu2024short}. Bold the best-performing model, and underlined indicates the second-best. We use a baseline GPT baseline (Section 5) and modify intermediate embeddings by imposing a hierarchical structure. Non-transformer-based, modified attention-based or hybrid architectures are not reported. }
\setlength{\tabcolsep}{2pt} 
\renewcommand{\arraystretch}{0.9} 
\begin{tabular}{llccc}
\toprule
\textbf{Attention Based Models} & \textbf{ListOps} & \textbf{Text} & \textbf{Image} \\
\midrule
Transformer \citep{vaswani2017attention} & 36.37 & 64.27 & 42.44  \\
Local Attention \citep{Tay2020b} & 15.82 & 63.98 & 41.46  \\
Linear Trans. \citep{vyas2020transformers} & 16.13 & \underline{65.90} & 42.34 \\
Linformer \citep{wang2020linformer} & 35.70 & 53.94 & 38.56 \\
Sparse Trans. \citep{child2019sparse} & 17.07 & 63.58 & 44.24  \\
Performer \citep{kaiser2020performer} & 18.01 & 65.40 & 42.77 \\
Sinkhorn \citep{tay2020sinkhorn} & 33.67 & 61.20 & 41.23 \\
Longformer \citep{beltagy2020longformer} & 35.63 & 64.02 & 40.83 \\
BigBird \citep{zaheer2020bigbird} & 36.05 & 64.02 & 40.83 \\
Luna-256 \citep{Ma2021Luna} & 37.25 & 65.78 & 47.86 \\
Reformer \citep{kitaev2020reformer} & 37.27 & 56.10 & 38.07 \\
\midrule
Non-Causal \\
FNET \citep{lee-thorp-etal-2022-fnet} & 37.27 & 56.10 & 38.07 \\
WavSPA \citep{zhuang2024wavspa} & \underline{55.40} & \textbf{81.60} & \underline{55.58} \\
\midrule
(Ours) GPT Baseline & 41.65 & 65.32 & 49.81 \\
\textbf{ (Ours) WaveletGPT} & \textbf{57.5} & \underline{66.38} & \textbf{59.81} \\
\bottomrule
\end{tabular}
\vspace{-0.5cm}
\end{table}

\section{Long Range Arena Benchmarks} We adapt our architecture for Long-Range Arena (LRA) tasks \cite{tay2021long}, testing on long-range prediction across text, images, and mathematical expressions. These tasks evaluate the model's ability to handle similarity, structure, and reasoning over extended contexts. We focus on transformer-based architectures, as recently reported by \cite{liu2024short}, while other variants include state-space, hybrid models or tweaking attention mechanisms. For text, we perform binary classification on the IMDb review dataset \citep{maas-EtAl:2011:ACL-HLT2011} using byte-level data with a context length of 2048 to determine if a movie review is positive or negative. We use CIFAR-10 from the LRA benchmark for images, classifying sequences of 3072 pixels into one of ten categories. Lastly, we benchmark on Long ListOps, testing the model's ability to understand hierarchically structured data in extended contexts. In our task, we use a version of ListOps of sequence lengths of up to \texttt{2K} to test the ability to reason hierarchically while handling long contexts. The model needs to access all tokens and model the logical structure of the inputs to make a prediction. The task is a ten-way classification task and is considerably challenging. We use the setup provided by \cite{khalitov2022sparse} to extract the data and be uniform with other benchmarks. We use identical architecture for all three modalities, only modifying the embedding matrix to accommodate different tokenizers and output categories. Our baseline consists of a 6-layer causal Transformer decoder with a model dimension of 32 and a feed-forward dimension four times that of the embedding dimension. We extract the last token of the sequence as a 32-dimensional embedding for classification, followed by a dense layer with 2048 neurons and a final dense layer corresponding to the number of categories. The input goes through an embedding layer that converts discrete tokens into a 32-dimensional vector. The input vocabularies are 256 for text/image and 16 for ListOps. The context lengths are 2048, 3072, and 1999 tokens, with 2, 10, and 10 as output categories. In our modified architecture, we introduce our waveletGPT module between each decoder layer, retaining half of the embedding dimensions as they are. For the other half, we use non-learnable kernels, increasing the kernel size from 2, 4, and 8 to 512 linearly for dimensions 16 to 32 while maintaining the causality assumption. This introduces highways that hierarchically process data at each embedding and Transformer decoder layer without adding parameters, similar to our approach for pre-trained LLM. As shown in Table 1, we achieve notable gains across all three modalities, where even minor improvements are worth reporting. We significantly outperform non-causal signal processing based methods, such as \citep{zhuang2024wavspa}, with nearly 2\% improvement on ListOps and 4.5\% on a much smaller architecture: ours has 32 dimensions and six layers compared to 128 dimensions/eight layers. We limit our comparison method for fairness only with vanilla Transformer architectures. We also compare two non-casual architectures incorporating signal processing-based ideas: FNET and WavSPA. Compared to non-causal FNet, our model significantly outperformed all three LRA tasks, achieving 20\%  improvement on ListOps and Image and 10\% on text. The most notable gain is in the ListOps task, which involves modelling a hierarchical, tree-like structure of math operations, making our model particularly suitable. To the best of our knowledge \cite{liu2024short}, this is the best result by a simple attention-based Transformer on LRA tasks.

\section{Conclusion and Future Work} We showcase a powerful incorporation of a core signal processing idea, namely wavelets, into large language model pre-training. By imposing a multi-scale structure onto every intermediate embedding, we achieve the same performance 40-60\% faster than a baseline architecture without adding any parameters. We achieve a substantial performance boost if we train for the same number of steps.  Further we show strong gains in LRA benchmarks without adding any parameters by giving every next token prediction access to multi-scale embeddings in every decoder layer. Our method generalizes across three modalities: raw text, symbolic music, and raw audio, giving similar performance speedups on several input representations, namely raw audio samples, acoustic tokens, MIDI tokens, byte text, math expressions, BPE tokens for text, raw image pixels and characters. This shows it is generic enough for improving pre-training performance across datasets and input representations.
\section{Acknowledgment}
This work was supported by the Stanford Institute of Human-Centered AI (Stanford-HAI) through a Google Cloud computing grant for the academic year 2023. The author is thankful to Prof. Dan Jurafsky for feedback and helpful discussions on the problem and its impact on the field.

\bibliography{example_paper}
\bibliographystyle{icml2024}

\section*{Impact Statement}
This paper presents work whose goal is to advance the field of 
Machine Learning by enabling LLMs to be trained faster and pushing the performance of the smaller architectures.  This will not only push the performance of the smaller architectures, but also yield faster inference. There are many potential societal consequences 
of our work, none which we feel must be specifically highlighted here, as it shares the same consequences in terms of positive and negative aspects of LLMs and AI systems which can potentially surpass human intelligence in near future.  
\nocite{langley00}
\end{document}